# Off-Line, Multi-Detector Intensity Interferometers I: Theory

Aviv Ofir[1], Astronomy and Astrophysics Dept, Tel Aviv University, Tel Aviv 69978 Israel, and Erez N. Ribak, Physics Dept., Technion, Haifa 32000, Israel

ABSTRACT

Stellar amplitude interferometry is limited by the need to have optical distances fixed and known to a fraction of the wavelength. We suggest reviving intensity interferometry, which requires hardware which is many orders of magnitude less accurate, at the cost of more limited sensitivity. We present an algorithm to use the very high redundancy of a uniform linear array to increase the sensitivity of the instrument by more than a hundredfold. When using an array of 100 elements, each almost 100m in diameter, and conservative technological improvements, we can achieve a limiting magnitude of about $m_b$=14.4. Digitization, storage, and off-line processing of all the data will also enable interferometric image reconstruction from a single observation run, and application of various algorithms at any later time. Coronagraphy, selectively suppressing only the large scale structure of the source, can be achieved by specific aperture shapes. We conclude that after three decades of abandonment optical intensity interferometry deserves another review.

Subject Heading: instrumentation: interferometers - instrumentation: high angular resolution - techniques: interferometric

1. INTRODUCTION

Amplitude (or Michelson) interferometry is today's mainstream technique for high angular resolution astronomy. Amplitude interferometers add the complex amplitude of the electromagnetic waves from two or more separate locations to produce a high-resolution brightness distribution, or image, of the source. Intensity interferometry, on the other hand, "interferes" the intensities of the electromagnetic wave via the correlation of the electrical currents generated by the detectors of the already-detected intensities. The main advantage of intensity interferometry is its mechanical robustness: the required opto-mechanical accuracy depends on the electrical bandwidth of the detectors and not on the wavelength of the light, and thus the mechanical precision required is relaxed by many orders of magnitude. This low path-length sensitivity also means that the existence of an atmosphere does not influence the performance of the instrument. The main disadvantages of intensity interferometry, which led to its demise, are its very low intrinsic sensitivity and the fact that the classical, two-detector intensity interferometer can not reconstruct the phase of the complex degree of coherence, and thus cannot be used to produce true images [1].

Gamo [2] proposed and Sato *et al* [3] proved experimentally that the three-detector intensity interferometer, which correlates the intensities from three separate detectors, can reconstruct the phase of the complex degree of coherence. Later, more algorithms to reconstruct the missing phase of the complex degree of coherence from amplitude-only measurements were proposed, by using second-, third-, and forth order intensity correlations [4], triple correlations and bispectra [5, 6 or 7], fractional triple correlation [8], and even by using just the usual second-order correlations and the Cauchy–Riemann equations [9].

Development of the two-detector intensity interferometer started in radio astronomy [10], but was expanded to the optical regime [11, 12] to culminate in the measurement of the angular diameter of 32 stars, during the operation of the Narrabri Stellar Intensity Interferometer (NSII) from 1965 to 1972. The low sensitivity prohibited observing stars fainter than $m_b$=2.5 though NSII used a pair of 30m$^2$ reflectors [1]. In fields other than astronomy intensity interferometry has been applied to nuclear physics (usually called "HBT effect") [13], ultra short laser pulses [14], characterization of the synchrotron radiation [15], hard disks head-disk spacing measurement [16], and measurement of electron temperature fluctuations in fusion plasmas [17]. It even seems that Fluorescence Correlation Spectroscopy (FCS), a technique regularly used in biology and chemistry, is actually intensity interferometry in disguise, and this similarity is especially striking when comparing works on FCS [18] and triple correlation [5,6]. The considerations in this paper apply not just in astronomy but wherever intensity interferometry is used.

---

[1] Address for correspondence: avivofir@wise.tau.ac.il

Fontana [19] generalized intensity interferometry to *N* detectors correlating all *N* currents to form a single output. We adopt Fontana's notations, specifically the multiple correlation function $F^{(N)}(\tau_1, \tau_2, ..., \tau_N) = \int_{-\infty}^{\infty} \langle I_1(t-\tau_1) \cdots I_N(t-\tau_N) \rangle dt$ and its excess above the zero-coherence term $\int_{-\infty}^{\infty} \langle I_1(t-\tau_1) \rangle \cdots \langle I_N(t-\tau_N) \rangle dt$ which will be designated $\Delta F^{(N)}$. As Fontana, we notate the first order correlation function as $g_j(\tau)$, where $\tau_i$ are the electrical delays added to each beam. We note that Fontana correlated *all* currents to form a *single* output, so Fig. 1 in [19] is somewhat misleading.

## 2. REDUNDANCY TO INCREASE SNR

### 2.1. High Redundancy of the Uniform Array

Firstly we describe the proposed instrument: we define reflectors as the surfaces of light collection and detectors as the series of the light-detecting instruments observing a *single* source. Mounted on each reflector there may be several detectors, each observing a different source where all the *j* detectors onboard each of the *N* reflectors point at the same source. In contrast with Fontana's *N*-detector intensity interferometer we record all signals directly after the amplifiers, and perform all correlations off-line, by software (Fig. 1). This setup will make it easier for us to use each signal many times, and to perform all other algorithms on the data at any later time.

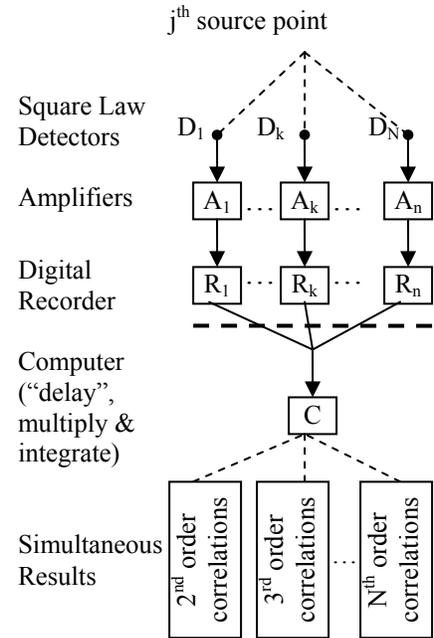

**Figure 1**: In our intensity interferometer the correlations are performed off-line, by software.

We will now show that for a linear array of many detectors with a uniform spacing *d*, this high redundancy can be used to effectively increase the overall signal to noise ratio (SNR). In general, one can compute not just the second order intensity correlation between detectors *a* and *b*, $F^{(2)}(\tau_a, \tau_b)$, but also the *m*th order correlation, $F^{(m)}(\{m\})$, of any *m*-sized subgroup of detectors $\{m\}$. The signal from each higher order of multi correlation $F^{(m)}(\{m\})$ will grow smaller with *m*. Still, the very high redundancy of $F^{(m)}(\{m\})$ will sometimes more than compensate for this. There are two types of redundancy: translational symmetry and high order expressions:

(1) From translational symmetry, one can see that any pair $(a,b)$ of detectors with a baseline of $|a-b|d$ is identical to any other pair of the same separation, and there are $N-|a-b|$ such pairs. Unlike amplitude interferometry, they can be added directly, since all phase information is already lost after detection.

(2) In the case of fields that obey Gaussian statistics, like stellar light, all high moments of the multi-correlation function (3rd and higher) can be expressed as a function of the first- and second- order correlations. This means that:

   a. The analytical expression for the correlation of *m* detectors – *a*, *b* and (*m*-2) other detectors – is the *m*th-order (*m*>2) correlation $F^{(2)}(\tau_a, \tau_b, \cdots, \tau_m)$, and it can be expressed as a function of low order correlations that will also include the specific expression $F^{(2)}(\tau_a, \tau_b)$.

   b. Reversing this relation, from each new subgroup one can construct a new expression of $F^{(2)}(\tau_a, \tau_b)$ by using the new $F^{(2)}(\tau_a, \tau_b, \cdots, \tau_m)$.

   c. Since there are $G = \binom{N-2}{m-2}$ subgroups of *N* with *m* members of which two are exactly *a* and *b*, there are also *G* different expressions for $F^{(2)}(\tau_a, \tau_b)$ in all subgroups of *m* members of *N* detectors.

d. Since the number of subgroups of *m* elements of *N* peaks at *m=N/2* while the signal from every higher order correlation is increasingly smaller, there is no gain in continuing beyond *m=N/2*.

Combining (1) and (2), the number of expressions of $F^{(2)}(\tau_a, \tau_b)$ possible with all $F^{(m)}(\{m\})$, or the redundancy of $F^{(2)}(\tau_a, \tau_b)$ in all subgroups of *N* is:

$$\sum_{m=2}^{\lfloor N/2 \rfloor} \binom{N-2}{m-2}(N-|a-b|) \qquad (1)$$

Thus giving a very high incentive to increase the number of reflectors *N*.

### 2.2. Signal

Before calculating the benefits of the high redundancy in terms of signal-to-noise ratio (SNR), we first need to generalize the procedure of writing the correlation functions *F*, now with two groups of variables {$m_1$} and {$m_2$} where any delay $\tau_i$ can be on either one or both {$m_1$} and {$m_2$}. We generalize Fontana's explanation of how to write $F^{(|m_1|+|m_2|)}(\{m_1\},\{m_2\})$ (after Eq. (29) in [19]): The allowed combinations of first order correlation functions can be deduced by representing all the distinct variables $\tau_i$ in $\{m_1\} \cup \{m_2\}$ as points, and every first order correlation function $g(\tau_j - \tau_i)$ as a line section between points *i* and *j*. The multiple correlation $F^{(|m_1|+|m_2|)}(\{m_1\},\{m_2\})$ is obtained by writing all possible ways to connect all the points with a continuous line beginning and ending at the same point, |$m_1$| + |$m_2$| sections long, and passing through each point the total number of times it appears in {$m_1$} and {$m_2$} (once or twice, in our case). Note that if some variables do appear more than once, then a term can appear $g(\tau_j - \tau_i) = g(0)$ to some power, which did not exist in [19]. When {$m_1$} ∩ {$m_2$} = {} this generalization reduces to Fontana's usual correlation function *F* with |$m_1$| + |$m_2$| variables (Fig. 2). We used short hand to write $F^{(m)}(\tau_a, \tau_b, , , \tau_m)$ as $F_{12...m}$.

Since the different subgroups of the array are partially overlapping, they are not statistically independent. We can correct for this statistical dependences between all the different representations of $F^{(2)}(\tau_a, \tau_b)$ by subtracting the cross-correlation of any new subgroup with all previous subgroups. This cross-correlation can be expressed as

$$cor(F_{\{m_1\}}, F_{\{m_2\}}) = \frac{\langle F_{\{m_1\}} \cdot F_{\{m_2\}}\rangle - \langle F_{\{m_1\}}\rangle\langle F_{\{m_2\}}\rangle}{\sigma(F_{\{m_1\}}) \cdot \sigma(F_{\{m_2\}})} = \frac{\langle F_{\{m_1\},\{m_2\}}\rangle - \langle F_{\{m_1\}}\rangle\langle F_{\{m_2\}}\rangle}{\sqrt{\langle F_{\{m_1\},\{m_1\}}\rangle - \langle F_{\{m_1\}}\rangle^2} \cdot \sqrt{\langle F_{\{m_2\},\{m_2\}}\rangle - \langle F_{\{m_2\}}\rangle^2}}, \qquad (2)$$

where $\sigma$ stands for standard deviation. After subtracting all multiply-counted representations, a true statistical meaning is established to the redundancy of the desired quantity, $F^{(2)}(\tau_a, \tau_b)$ in our example.

Let us relate these results to some real world values. The two-detector intensity interferometer has a signal [1]

$$Signal^{(2)} = e^2 b_\nu A_1 A_2 |\gamma_{12}(0)|^2 \int_0^\infty \alpha^2(\nu) n^2(\nu) d\nu \qquad (3)$$

Where *e* is the electron's electrical charge, $b_\nu$ is the detector's electrical bandwidth, $A_1$, $A_2$ are the reflector's areas, $\alpha$ is the detectors' quantum efficiency at frequency $\nu$ (which are assumed to be equal), and *n* is the photon flux density at $\nu$. Changing to Fontana's notation and generalizing for an *m*-detector intensity interferometer subgroup, each subgroup will create a signal of

$$Signal^{(m)} = e^m b_\nu (A_1 \cdots A_m) \cdot \Delta F^{(m)}(\tau_1,...,\tau_m) \cdot \int_0^\infty \alpha^m(\nu) n^m(\nu) d\nu \qquad (4)$$

In our specific case of a linear, uniformly-spaced array this signal will be enhanced by the increased statistical significance found in the many representations of $F^{(m)}$ (Eq. [1]), after correcting for multiply-counted representations.

$$F^{(3+2)}\left(\{\tau_1,\tau_2,\tau_3\},\{\tau_3,\tau_4\}\right) = F_{12334} =$$
$$= 2F_{234}F_{13} + 2F_{134}F_{23} + 2F_{123}F_{34} \qquad + \qquad 2F_{1234}g(0)$$

**Figure 2** - The expression for the fifth order $F_{12334}$: Every line segment between two points is a first-order correlation function between the two corresponding detectors. A thick line means two passes along the same section [eg. $F_{12}=g^2(\tau_1-\tau_2)$]. The circling arrow means $g(\tau_i - \tau_i) = g(0)$. Functions $F$ were used to denote the five first-order correlation in each of the above terms (drawings). The multiplicity of each term (two in the above example) is the number of possibilities to re-order the different sections and still get the same final continuous line beginning and ending at point 1.

2.3. Noise

The noise in an optical intensity interferometer is dominated by "shot noise", cause by the discreteness of electrical charges [11]. To calculate the noise in a shot-noise dominated environment one only needs the very first order of the different intensities in the subgroup, so the expression for the noise (squared) for the two element array can be well approximated simply by [11]

$$(Noise)^{2\,(2)} = \langle I_1(t-\tau_1)\rangle \cdots \langle I_m(t-\tau_m)\rangle . \tag{5}$$

We fused Eq. (5) and a result by Mandel [20] to a format similar to Eq. (3):

$$(Noise)^{2\,(2)} = 2e^2 A_1 A_2 \left(\frac{b_\nu}{T_0}\right) \int_0^\infty \alpha^2(\nu)n^2(\nu)d\nu . \tag{6}$$

This is easily generalized to an *m*-element subgroup of an intensity interferometry array

$$(Noise)^{2\,(m)} = 2e^m (A_1 \cdots A_m) \left(\frac{b_\nu}{T_0}\right) \int_0^\infty \alpha^m(\nu)n^m(\nu)d\nu , \tag{7}$$

Which will be applied to every new subgroup of *N* which is been considered (pairs, triplets, etc.).

2.4. SNR Calculation Algorithm

Now that we have both signal and noise for all subgroups, we give an algorithm to calculate the SNR of a complete *N*-detector intensity interferometer using the high redundancy of all its subgroups. For the 1*d* baseline signal $F^{(2)}(\tau_1,\tau_2)$ in an *N*-detector array (capital letters stand for accumulating variables)

1. For all subgroups $\{m\} = \{\tau_1,\tau_2,\tau,\tau\cdots\}$ which contain $(\tau_1,\tau_2)$ or translationally symmetric pairs (all subgroups with 1*d* baseline).
   1.1. Add signal from $\{m\}$ to SIGNAL(1*d*)
   1.2. Subtract all correlations of $\{m\}$ with previously used $\{m\}$s from SIGNAL(1*d*).
   1.3. Add (noise)$^2$ from $\{m\}$ to NOISE$^2$(1*d*)
   1.4. Next $\{m\}$
2. SNR of 1*d* = (SIGNAL) / root (NOISE$^2$)
3. Next *F* function (next separation or next multiple correlation) from step 1.

Note that if the optical bandwidth is narrow enough so that both $\alpha$ and *n* are constant for all relevant $\nu$, the resultant SNR of any subgroup of *m* detectors is proportional to

$$SNR \propto (A\alpha n)^{m/2}. \tag{8}$$

## 3. SIMULATIONS AND PROJECTED CAPABILITIES

### 3.1. Approximation and Comparison Base

In order to check our algorithm for some general object, we approximated $g$, the first order correlation function between any two detectors. Since $0 \leq |g| \leq 1$ we will take it to be $g = 1/2$ for all pairs, since a properly chosen detector spacing $d$ should achieve $\langle |g| \rangle \approx 1/2$ for maximum dynamic range (photon anti-bunching experiments may also yield $g \approx -1/2$). We therefore substitute $g = 1/2$ in Fontana's result, so for the $j^{th}$ source the signal part of the output of the intensity interferometer, the excess correlation $\Delta F$ is:

$$\Delta F^{(N)} \approx \frac{\prod_{i=1}^{N} \sigma_i}{(2\pi)^{N-1}(c\varepsilon_0)^N} \sum_j U_j \frac{(N-1)!}{2} \tag{9}$$

Inserting Eq. (9) into Eq. (2) gives the approximated correlation between any two subgroups of $N$. Since $(2m_1-1)!/2 \gg ((m_1-1)!/2)^2$ already at small $m_1$, in our simulation we used:

$$cor(F_{\{m_1\}}, F_{\{m_2\}}) \approx \frac{(m_1+m_2-1)! - (m_1-1)!(m_2-1)!/2}{\sqrt{(2m_1-1)!}\sqrt{(2m_2-1)!}} \tag{10}$$

for the statistical correlation of any one subgroup of size $m_2$ with a previously used subgroup of size $m_1$. To compute real-world results we used the NSII performance figures [1]: wavelength $\lambda = 438.4$nm, electrical bandwidth $b_\nu = 100$ MHz, quantum efficiency $\alpha = 0.2$, system efficiency $\Sigma = 0.2$, and reflectors area $A = 30$ m$^2$, to achieve SNR = 27 for integration time $T_0 = 1$hr of a star of blue band magnitude $m_b = 0$ (i.e. a photon flux density of $n = 5 \cdot 10^{-5}$ photons m$^{-2}$ sec$^{-1}$ Hz$^{-1}$). The fact that the technological parameters scaling laws are experimentally verified will also allow us to correctly allow for all technological improvements since 1972.

### 3.2. Results and Analysis

Since we know that the redundancy of $\{m\}$ in Eq. (9) is highly dependant on $N$, we investigated the effect of the quantities in question. Since the quantum efficiency $\alpha$ has a relatively narrow range to change, we continue and change $A \cdot n$ (namely the photon flux density at each detector). In some non-astronomical applications $n$ is controllable, and increasing it will give similar results to increasing $A$, since what matters is the product $A\alpha n$. In astronomy $n$ is uncontrolled, and Eq. (8) means a strong incentive to choosing the wavelength in which $n$ is maximal for each source. We will therefore use the number of reflectors $N$ and the area of the single reflector $A$ as the main variables in our analysis (see Section 4 for discussion on the case when apertures $A$ can no longer be considered "small").

In Figure 3 we plotted the SNR of the correlation function for 1$d$ separation of several offline, multi-detector, linear and uniform intensity interferometers, each with a different (but uniform) reflector area $A$, as a function of the number of reflectors in the array $N$. The individual reflectors' area starts at 30m$^2$ (as in NSII) and double the effective linear size (quadruple $A$) at each new plot up to an area of 7680m$^2$, or a single reflector diameter of ~100m (similar to current Extremely Large Telescope concepts, but our reflectors are crude light buckets and not telescopes). A clear change in behavior is evident on the 7680m$^2$ plot around $N \approx 10$-20, and a similar, more subtle, change can be seen on the 1920m$^2$ plot (near $N \approx 50$-60). These plots *do not* illustrate technological dependence, being taken to be the same as those of NSII.

The leftmost point ($N = 2$) on the 30m$^2$ (bottom) plot is the known NSII performance quoted at the end of §3.1. The uniform (on log scale) spacing between all the left-most points of each plots (all $N$=2) demonstrates the known linear scaling law of the two-detector intensity interferometer with respect to reflectors' area [1].

The entire 30m$^2$ plot illustrates the 4.91 fold improvement (over the 2-element instrument) of the SNR of when we simulated a 100-element intensity interferometer, each of them identical to the ones used by NSII. This curve is entirely the result of the translational redundancy symmetries of different pairs - there are no observable differences if one ignores all higher order contributions.

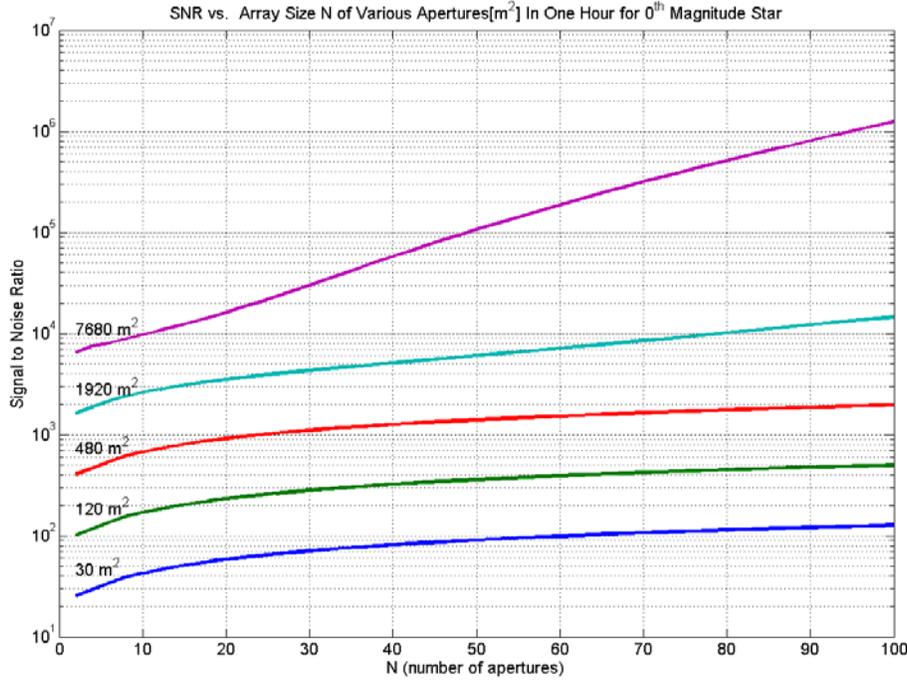

**Figure 3**. The SNR of different arrays, all using NSII technology, each with a different aperture as indicated, vs. N for a separation $|a_1-a_2|=1$ in one hour for a star of 0 magnitude. Apertures range from 30m² (as in NSII) to 7680 m². A clear change in behavior is seen on the 7680m² plot, and a similar, more subtle, change can be seen on the 1920 m² plot (around N=54).

Let us now explain the shape of the upper plots of Figure 3: The overall behavior of all the plots with respect to *N* is tapering down with increasing *N* – the translational symmetry behavior. Apart from this behavior, the change in behavior of the 7680m² plot implies that a new element becomes important around $N \approx 10 - 20$. Since we already know that the 30m² plot is entirely the product of the usual second order correlations (no contribution from high orders) we shall call that change in behavior a transition from "two-correlation regime" to "multi-correlation regime". We will now find the condition in which the contribution of all pairs is equal to that of all next-level correlations, i.e., all triplets. In §2.1 we showed that the redundancy of all subgroups of size *m* is $\binom{N-2}{m-2}(N-|a_1-a_2|)$ while signal from each of these subgroups is proportional to $(A\alpha n)^m$, giving a total signal from all subgroups of certain size *m*:

$$\binom{N-2}{m-2}(N-|a_1-a_2|)(A\alpha n)^m \tag{11}$$

Comparing this expression for *m*=2 (pairs) and *m*=3 (triplets) will give us an *N*-A relation determining when one should see the contribution from all triplets equal to that from the pairs. For the separation of $|a_1 - a_2|=1$ depicted in Figure 3, we get $(N-2)A\alpha n = 1$ which means that an array with reflector size *A*=7680m² and a 0 magnitude star will be dominated by triplets when *N* = 15.02, or alternatively, that an array of 15 detectors will be triplets-dominated for reflector sizes of 7692m² and up. This procedure can be applied to also check when the quadruples start to contribute even more than the triplets, which happens at *N* = 29, and quintuplets will contribute more than the quadruplets at *N* = 43, sextuplets will dominate at *N* = 57, septuplets at *N* = 71, octuplets at *N* = 85 and finally nonuplets at *N* = 99. The end result is such a long exponential rise because it is actually the stacking of all the above contributions. Similarly, a transition to triplets domination, although not as pronounced, can be observed also for the *A* = 1920m² around *N* = 54. Now we can explain why no such transition has been observed at the lower area plots, like the 30m² plot, as the transition point for triplets domination for it is at $N = 3335$, and for the 480m$^2$ plot the transition point is at $N = 210$.

This behavior is almost completely technology-independent, but the absolute values are very much effected by technology: an estimate for the technological improvements since 1972 give, with the scaling laws given in [1], a 40 fold improvement by conservatively changing $b_v$ to 1GHz, $\alpha$ to 0.8, $\Sigma$ to 0.8, and still only one optical channel, $p=1$. We choose not to pursue the technological options further here, but we note that a measure to the conservatism in our estimate is the 1969 paper by Twiss arguing that technology alone could increase the SNR for the two-detector intensity interferometer by a factor of 80 [21].

Dimmer sources affect the result in a way similar to smaller reflectors since SNR of any subgroup is proportional to $(A\alpha n)^m$. Following Hanbury Brown and Twiss, we define the limiting magnitude of the instrument as the magnitude where we only get SNR of 3 after one hour of integration, then the limiting magnitude of the 7680m$^2$, 100 element off-line intensity interferometer is slightly more than the 10$^{th}$ magnitude using the NSII technology, and about 14.4 magnitudes when the above mentioned conservative technological improvements are considered. At that point one will notice that: (i) Redundancy from translational symmetry do not depend on the source's strength, so this effect remains and contributes (see the behavior of the lower plots of Figure 3). (ii) Redundancy from higher order correlations is highly dependant on the photon flux $n$, and its contribution is negligible for sources dimmer than magnitude 3 for the ~100m diameter reflectors (using NSII technology). Virtually all the instrument's capabilities beyond this point are due to the shear area of the reflectors and the translational symmetry. (iii) In calculating the preceding figures were did not include the coronagraphic effect (see §4) as it depends also on the source's angular size and the reflector's shape, and can thus be chosen to have a modest impact.

## 4. CORONAGRAPHY WITH LARGE APERTURES

In what appears to be in some conflict with our computations here, the SNR cannot be indefinitely increased simply by increasing the reflector size $A$. Intensity interferometry is based upon the assumption that the source is a "point" source (i.e., smaller than the diffraction limit of a single reflector). By increasing the reflectors to very large diameters one realizes that some stars can no longer be considered as point sources. This effect was accounted for by Hanbury Brown and Twiss by introducing the partial coherence factor $\Delta(v)$ [12] which reduces the observed correlation for partially resolved sources, cancels the observed correlation altogether for completely resolves sources, and complicates the interpretation considerably as $\Delta(v)$ also depends on the size and shape of the source.

Yet, we foresee a way to utilize that effect to our advantage for searching and characterizing extremely high dynamic range objects, like binaries, multiples and even extra-solar planets: when one observes an extra-solar planetary system around sun-like stars one notices three length scales: the orbital distances of the planets, the size of the star and the sizes of planets. If we choose a reflector size between the size needed to resolve the star and the planets, we would find that $\Delta(v)$ has already significantly reduced the stellar signal, but it has yet to affect the planetary signals. By "using" the partial coherence factor we can selectively attenuate the signal from all large scale structures of the source (which are almost always the brighter structures), and don't need dynamic range as wide as before, which means that our instrument is now also a coronagraph. This quality of the large-aperture intensity interferometers enables one to apply coronagraphy to stars other than the Sun, and to do it from the ground. We clarify that this effect will reduce the signal from object scales close to- and larger than- the diffraction limit of the single dish, and not from object scales close to the diffraction limit of the baseline.

For example, for a 7680m$^2$ square shaped reflector observing a star like the sun at a distance of 10pc, the partial coherence factor of 0.72 reduces the pair-wise correlation (stellar signal) by 28%. This modest attenuation can be enhanced by considering elongated or rectangular reflectors. It doesn't matter which side is longer as long as the round symmetry of the object is not broken. For the same reflector area, this setup will reduce the pair-correlation stellar signal by a factor of ~2.5 for an aspect ratio of 1:4 in the reflector. From that behavior of the partial coherence factor, one concludes that there must be two extreme reflector shapes: one which minimizes the partial coherence factor (and thus maximizes the coronagraphic effect) and one which does the opposite – maximizes the partial coherence factor (and minimizes the coronagraphic effect). These shapes depend on the source's shape, but can be computed for a uniform, circular source by variational calculus by minimizing (maximizing) the expressions for $\Delta(v)$ given in Appendix 3 of [12] for a two-detector intensity interferometer. One then must ask what will happen to the third and higher-order correlations when large reflectors are used. We only made initial calculations which seem to indicate that the multi-aperture coronagraphic effect is far more pronounced, perhaps by orders of magnitude, compared to that obtained with two apertures. Renewed interest in intensity interferometry might justify additional development of this subject.

5. CONCLUSIONS

We presented an algorithm for the improvement of the SNR of an evenly spaced off-line multi-detector intensity interferometer by utilizing its very high redundancy. We showed that by stacking many contributions in the multi-correlation regime the SNR of such an array scales approximately exponentially with ($NA\alpha n$) (fig. 3 top curve). We demonstrated the algorithm on the simplest term $F^{(2)}(\tau_1, \tau_2)$ but the generalization to triple and higher correlation is straightforward. We showed that translational symmetry improves the performance of the instrument by a factor of about five, and that multi correlation can further improve that performance significantly (a total improvement of more than 190-fold), under the investigated conditions. This improvement is made possible by the offline processing of the data that allows us to "use" each photon several times and thus to alleviate the low intrinsic sensitivity of intensity interferometers, to achieve a limiting magnitude of about 14.4 magnitudes, when using 100-element, 7680m$^2$ each, conservatively technologically improved array. Indeed, off-line processing of the data enables to reconstruct the whole complex correlation function (in $N$-1 points) from a single observation run by using all available $F^{(m)}(\{m\})$. Since the number of detectors $N$ is expected to be at least few dozens, the ($u, v$) coverage will be good enough to reconstruct an optical interferometric image with resolution in the μas range (100 elements, each 100m in diameter means a *minimum* baseline of 10Km) without having to fit the visibility curve to some model. In a parallel paper [22] we discuss further implications and uses of the proposed instrument, which greatly enhance the scientific productivity of the instrument..

When intensity interferometry was first introduced it was believed that all phase information is lost, and one cannot hope to reconstruct a true image using intensity interferometry. Today there exist many different algorithms to reconstruct both amplitude and phase information. In this context we presented our algorithm for the improvement of the SNR, and believe that this algorithm is not the only one possible. An intensity interferometry array which can record all information and process it later on will allow the application of any new algorithm to all previous observations.

Contemporary astronomy is plagued by the need to have optical surfaces smooth and distances fixed to a fraction of the wavelength. Multi-detector optical intensity interferometry offers a way out of this restriction, even if not for the faintest of objects. After 35 years, results obtained with intensity interferometry are still the state of the art in terms of resolution and wave length. The main drawback of intensity interferometry is sensitivity, but using the above hardware and software improvements and scaling laws one understands that a multi-detector array could be used as a present day technique answering present day questions, and indeed deserves another review.

This work is based upon a master's thesis by A. Ofir [23] who can be reached at avivofir@wise.tau.ac.il.. Parts of this work were supported by the European Interferometry Initiative through OPTICON (an EU Framework VI program).